 \documentclass[final,5p,times,twocolumn]{elsarticle}

\usepackage{amssymb}

\journal{Physica B}

\begin{document}

\begin{frontmatter}



\title{Phonon broadening from supercell lattice dynamics: random and correlated disorder}



\author[ox,di]{A.~R.~Overy}
\author[ox]{A.~Simonov}
\author[di]{P.~A.~Chater}
\author[or]{M.~G.~Tucker}
\author[ox]{A.~L.~Goodwin}

\address[ox]{Department of Chemistry, University of Oxford, Inorganic Chemistry Laboratory, South Parks Road, Oxford OX1 3QR, U.K.}
\address[di]{Diamond Light Source, Chilton, Oxfordshire, OX11 0DE, U.K.}
\address[or]{Oak Ridge National Laboratory, 1 Bethel Valley Road, Oak Ridge, Tennessee, USA.}

\begin{abstract}

We demonstrate how supercell implementations of conventional lattice dynamical calculations can be used to determine the extent and nature of disorder-induced broadening in the phonon dispersion spectrum of disordered crystalline materials. The approach taken relies on band unfolding, and is first benchmarked against virtual crystal approximation phonon calculations. The different effects of mass and interaction disorder on the phonon broadening  are then presented, focussing on the example of a simple cubic binary alloy. For the mass disorder example, the effect of introducing correlated disorder is also explored by varying the fraction of homoatomic and heteroatomic neighbours. Systematic progression in the degree of phonon broadening, on the one hand, and the form of the phonon dispersion curves from primitive to face-centered cubic type, on the other hand, is observed as homoatomic neighbours are disfavoured. The implications for rationalising selection rule violations in disordered materials and for using inelastic neutron scattering measurements as a means of characterising disorder are discussed.

\end{abstract}


\end{frontmatter}


\section{Introduction}

Phonon broadening plays a key role in thermal transport behaviour and so is of central significance in the design of functional materials whose performance depends on thermal conductivity \cite{Toberer_2012, Cahill_1988}. In the topical field of thermoelectrics, for example, low thermal conductivity is desirable as it helps maintain a thermal gradient during current flow \cite{Snyder_2008, Yang_2016, Zeier_2016}. Indeed, the key figure of merit for thermoelectrics ($zT$) is inversely proportional to thermal conductivity. Because broadening reduces phonon lifetimes---and hence the lattice contribution to thermal conductivity---there is significant interest in understanding and controlling microscopic mechanisms by which it might be controlled \cite{Delaire_2011, Zhao_2014}.

There are two principal effects responsible for phonon broadening. First, anharmonic phonon--phonon coupling increases phonon scattering rates and so, in turn, reduces phonon lifetimes; the involvement of heavy atoms can further amplify the contribution at low energies, where the effect on thermal conductivity is usually greatest. Anharmonic broadening is often strong in systems with incipient ferroelectric instabilities and/or with significant electron--phonon coupling \cite{Delaire_2011}. This is the mechanism generally thought relevant to thermoelectrics such as PbTe \cite{Delaire_2011}, and is often accompanied by a dramatic signature in the phonon dispersion where a subset of phonons disperse in energy rather than wave-vector (so-called `waterfall' phonons \cite{Hlinka_2003}). A second origin of phonon broadening is that of static disorder. We use the term `static' to describe structural perturbations that are long-lived with respect to phonon lifetimes: a clear example is compositional disorder, but we also include long-lived electronic-driven distortions such as Jahn Teller (JT) instabilities, charge localisation, and polaron§ formation (where appropriate). The exploitation of disorder to moderate thermal conductivity is a common strategy in thermoelectric design; a topical example is the family of alloys based on the Heusler and half-Heusler structure types \cite{Downie_2013, Asaad_2016,  Xie_2014, Fu_2015}. The notoriously poor thermal conductivity of quasicrystal phases might also be understood (in general terms) in this same context \cite{Pope_2004, Courtens_2000}.

Static disorder need not be random, and may instead be associated with a (perhaps very) specific set of periodicities---this is the case of so-called `strongly correlated' disorder \cite{Keen_2015, Overy_2016} found in phases such as water ice \cite{Pauling_1935} and BaTiO$_3$ \cite{Megaw_1947, Senn_2016}. Since phonons also represent a correlated perturbation of the atomic positions in a crystal and are associated with well-defined periodicities \emph{via} the corresponding wave-vector $\mathbf k$, there is the potential for selective coupling between disorder and phonons. In other words, strongly correlated disorder allows in principle for targetted phonon broadening where only particular phonon branches---determined by the type of disorder---exhibit increased widths while others are unaffected by the presence of static disorder. We have suggested elsewhere \cite{Overy_2016} that this phenomenon may provide a mechanism of manipulating separately electronic and thermal conductivities as required by the `phonon--glass--electron--crystal' paradigm of thermoelectric design \cite{Snyder_2008}. In general, however, the extent and nature of disorder--phonon coupling for various types of correlated disorder remains relatively poorly understood, not least because of the axiomatic role of structural periodicity in conventional lattice dynamical theory \cite{Dove_1993}.

In this context, there is a clear need for straightforward lattice dynamical calculations capable of quantifying the effect of arbitrary types of static disorder on the phonon dispersion spectrum. A common and computationally inexpensive method for treating disorder within established harmonic lattice dynamical theory is \emph{via} the so-called `virtual crystal approximation' (VCA), a mean-field-like approach \cite{Abeles_1963}. The VCA determines the average effect of disorder on phonon dispersion and treats phonon broadening as a perturbation within the context of Klemens-Callaway theory \cite{Klemens_1955, Klemens_1957, Callaway_1959}. This theory holds well for low-frequency phonon modes and low levels of disorder, but breaks down in other scenarios \cite{Mattis_1958, Kamitakahara_1974}. At the other extreme of computational difficulty is the use of density functional theory (DFT) calculations to calculate excitation energies for explicit atomistic realisations of disordered states \cite{Alam_2011}. Such calculations are currently limited to systems of order $\sim10^2$ atoms, which is too small for generation of  phonon dispersion curves suitable for comparison to experimental data.

A compromise between these two extremes, accessible for systems where interactions may be effectively parameterised, is to carry out lattice dynamical calculations using disordered supercell configurations representing atomistic realisations of the particular state of interest. These calculations can be performed either in the (quasi-)harmonic limit or so as to account for anharmonic interactions using molecular dynamics (MD) simulations \cite{Larkin_2013}. In either case, off-the-shelf lattice dynamical tools can be used, such as the general utility lattice dynamical program (GULP) \cite{Gale_1997}. This approach has been used to explore the effects of \emph{e.g.}\ random substitution and force-constant disorder on phonon widths and eigenmode localisation in canonical alloys and related systems \cite{Larkin_2013, Beltukov_2013}.

In this paper we extend this approach to the calculation of full phonon dispersion relations for hypothetical disordered states on the primitive cubic lattice. In particular, we consider the case of non-random disorder and illustrate the gradual and continuous progression in phonon dispersion for a simple system as the strength of correlations is gradually increased. Our emphasis is on the ease of implementation of the supercell lattice dynamical (SCLD) approach detailed in Ref.~\cite{Overy_2016}, with a view to rapid exploration of the relationship between correlated disorder and phonon broadening for a variety of lattice types. Our paper is arranged as follows. We begin by summarising the methodology of Refs.~\citenum{Larkin_2013, Beltukov_2013}, our particular implementation with the GULP software, and detailing a benchmarking exercise within the VCA. In section \ref{results} we report the disorder-broadened phonon dispersion curves for a variety of systems based on the simple cubic lattice. We distinguish the effects of substitutional and force-constant disorder, and demonstrate the effect of including substitutional correlations in a random alloy on its corresponding phonon dispersion. We conclude with a discussion regarding the implications for rationalising selection rule violations in disordered materials and for using inelastic neutron scattering measurements as a means of characterising disorder.

\section{Methodology}\label{methods}

Conventional harmonic lattice dynamical calculations (\emph{e.g.}\ as implemented in the GULP software \cite{Gale_1997}) make use of the dynamical matrix $\mathbf D(\mathbf k)$ defined at wave-vector $\mathbf k$ in terms of the elements
\begin{eqnarray}
D_{\alpha j,\beta j^\prime}(\mathbf k)&=&\frac{1}{(m_jm_{j^\prime})^{1/2}}\sum_{\ell^\prime}\left[\Phi_{\alpha\beta}\left(\begin{array}{cc}j&j^\prime\\ 0&\ell^\prime\end{array}\right)\right.\\& &\left.\vphantom{\left(\begin{array}{cc}j&j^\prime\\ 0&\ell^\prime\end{array}\right)}\qquad\times\exp\{{\rm i}\mathbf k\cdot[\mathbf r(j^\prime\ell^\prime)-\mathbf r(j0)]\}\right].
\end{eqnarray}
Here, $\left(\begin{array}{c}j\\ \ell\end{array}\right)$ denotes the atom $j$ in unit cell $\ell$ with position $\mathbf r(j\ell)$, $\alpha,\beta\in\{x,y,z\}$, and $\Phi$ is the force-constant matrix
\begin{equation}
\Phi_{\alpha\beta}\left(\begin{array}{cc}j&j^\prime\\ 0&\ell^\prime\end{array}\right)=\frac{\partial^2W}{\partial u^{\alpha}_{j\ell}\,\partial u^{\beta}_{j^\prime\ell^\prime}}
\end{equation}
that relates variations in the lattice energy $W$ to displacements $u$. A well-known result of lattice dynamical theory \cite{Dove_1993} is that the $3Z$ eigenvalues of $\mathbf D(\mathbf k)$ are the squared normal mode frequencies $\omega^2(\mathbf k,\nu)$ (here $Z$ is the number of atoms in the unit cell and $\nu$ labels the mode) and the corresponding eigenvectors $\mathbf e(\mathbf k,\nu)$ describe the associated atomic displacement patterns. By evaluating $\mathbf D(\mathbf k)$ for a suitably dense set of wave-vectors $\mathbf k$ the phonon dispersion curves can be assembled.

The SCLD approach at the heart of our study differs in that a supercell containing $N$ conventional unit cells is used. We replace the original atom labels $j$ by the atom-cell labels $jl$ to track the position of the corresponding unit cell $l$ within the SCLD supercell. Each atom $jl$ within the supercell may now be assigned arbitrary mass $m(jl)$; likewise the force constants $\Phi_{\alpha\beta}\left(\begin{array}{cc}jl&j^\prime l^\prime\\ 0&\ell^\prime\end{array}\right)$ can vary throughout the configuration---\emph{i.e.}\ for different choices of $l,l^\prime$---even for constant $j,j^\prime,\ell^\prime$. The incorporation of disorder in $m$ or $\Phi$ usually means that the supercell configuration must first relaxed to minimise $W$ and to determine the equilibrium atomic coordinates. Whether or not relaxation is required, the corresponding dynamical matrix now contains a total of $3ZN\times3ZN$ entries.

Diagonalisation of $\mathbf D(\mathbf k=\mathbf 0)$ yields the $3ZN$ frequencies $\omega(\nu)$ and eigenvectors $\mathbf e(\nu)$ (each with $3ZN$ components $e_{jl}^{\alpha}(\nu)$, indexed by atom-cell $jl$ and cartesian direction $\alpha$) associated with usable normal modes of the supercell. We proceed by projecting these $3ZN$ modes onto a set of $N$ $\mathbf k$-points as allowed by the dimensions of the supercell. For isotropic supercells of length $L=\sqrt[3]{N}$, for example, the allowed $\mathbf k$-points (given in reduced form) are
\begin{equation}
\mathbf k=\left(\frac{n_1}{L},\frac{n_2}{L},\frac{n_3}{L}\right);\quad n_i\in\{0,1,\ldots,L-1\}.
\end{equation}
Hence finer $\mathbf k$-meshes require larger supercells. The projection $\rho$ for each $\mathbf k$ is determined for all $\omega$ by the relation
\begin{equation}
\rho(\mathbf k,\omega)=\sum_\nu\left\{\frac{\delta[\omega-\omega(\nu)]}{N}\sum_{jl,\alpha}\left|\sum_le_{jl}^{\alpha}(\nu)\exp[{\rm i}\mathbf k\cdot\mathbf r(l)]\right|^2\right\},
\end{equation}
where $\mathbf r(l)$ is the position vector for cell $l$ in the supercell. Our formalism here is analogous to the structure factor expression given in Eq.~(21) of Ref.~\cite{Beltukov_2013}. It is necessary to average over symmetry-equivalent $\mathbf k$ points, with the accuracy of $\rho(\mathbf k,\nu)$ being improved by further averaging over multiple independent configurations (we typically use five).


\begin{figure}
\begin{center}
\includegraphics[width=\columnwidth]{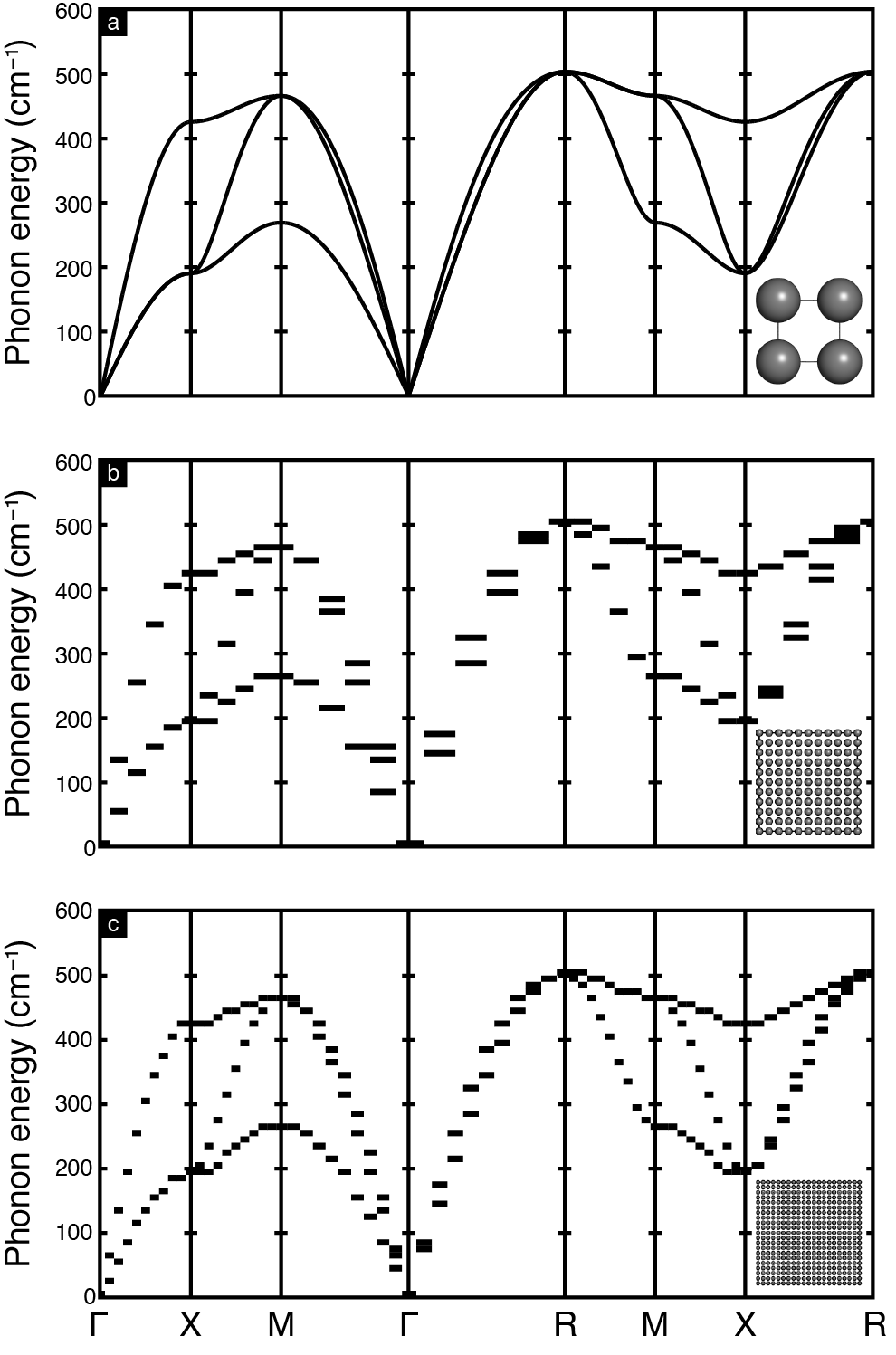}
\end{center}
\caption{Consistency test for SCLD calculations using a periodic model system. (a) Phonon dispersion curves determined using conventional dynamical matrix calculations as implemented in GULP (Ref.~\cite{Gale_1997}). The relevant parameterisation is described in the text and in Table~\ref{table1}. The simple cubic cell contains a single atom; hence there are three (sometimes degenerate) branches of the phonon dispersion at each $\mathbf k$-point. (b,c) The corresponding phonon dispersion curves determined using the SCLD approach described in the text for (b) $10\times10\times10$ and (c) $20\times20\times20$ supercells. Representations of the GULP unit cells are given in the bottom-right corner of each set of phonon dispersion curves \label{fig1}}
\end{figure}

\begin{table}[b]
\caption{ A parameter summary for the virtual crystal approximant described in the text.\label{table1}}
\begin{center}
\begin{tabular}{l|c}
Parameter & VCA value\\
\hline\hline
   $a$ / \AA & 10\\
$m$ / amu & 60\\
   $k_{\rm{harm}}$ / eV\AA $^{-2}$ & 10\\
   $r_0$ / \AA&10 \\
   $k_{\rm{angle}}$  / eV rad $^{-2}$ & 10\\
   $\theta_0$ / $^\circ$&90\\         
    \hline\hline
\end{tabular}
\end{center}
\end{table}

We demonstrate the consistency of this approach with conventional lattice dynamical calculations using a periodic simple cubic lattice. This example will later correspond to the VCA limit for a set of binary alloys that we consider. Fig.~\ref{fig1}(a) shows the harmonic phonon dispersion relations obtained using conventional lattice dynamical methods parameterised as follows. We consider a single ``average'' atom (mass $m$) in a primitive cubic unit cell (lattice parameter $a$) that interacts with its six nearest neighbours (separated by $\langle100\rangle$ lattice vectors) via effective harmonic bond stretching and bond bending terms:
\begin{eqnarray}
E_{\rm{harm}}&=&\frac{1}{2}k_{\rm{harm}}(r-r_0)^2;\\
E_{\rm{angle}}&=&\frac{1}{2}k_{\rm{angle}}(\theta-\theta_0)^2.
\end{eqnarray}
The actual parameters used are listed in Table~\ref{table1}; these have been chosen to ensure elastic stability and broad dispersion in phonon energies.

The corresponding SCLD phonon dispersion curves obtained using the same parameterisation for $10 \times 10 \times 10$ and $20 \times 20 \times 20$ supercells are shown in Fig.~\ref{fig1}(b,c). Because there is no disorder in these supercell configurations, then the projections $\rho(\mathbf k,\omega)$ take the discrete values 0 or 1 (strictly). The match between SCLD and conventional lattice dynamical calculations is exact, with the only difference between the two being the finite $\mathbf k$-grid on which the SCLD calculations are obtained.

\section{Results}\label{results}

Having established the viability of the SCLD approach in the mean-field limit of the VCA, we turn now to the key focus of our study: namely, determination of the intrinsic phonon broadening resulting from static disorder. Our implementation involves varying the masses $m(jl)$ and/or force-constants $\Phi_{\alpha\beta}\left(\begin{array}{cc}jl&j^\prime l^\prime\\ 0&\ell^\prime\end{array}\right)\equiv k_{\rm{harm}},k_{\rm{angle}}$ throughout a disordered configuration, as envisaged in Section~\ref{methods}. For simplicity, we focus on a single family of related systems: simple cubic binary (AB) alloys containing equal quantities of the two components.

\subsection*{Random disorder}

\begin{figure}
\begin{center}
\includegraphics[width=\columnwidth]{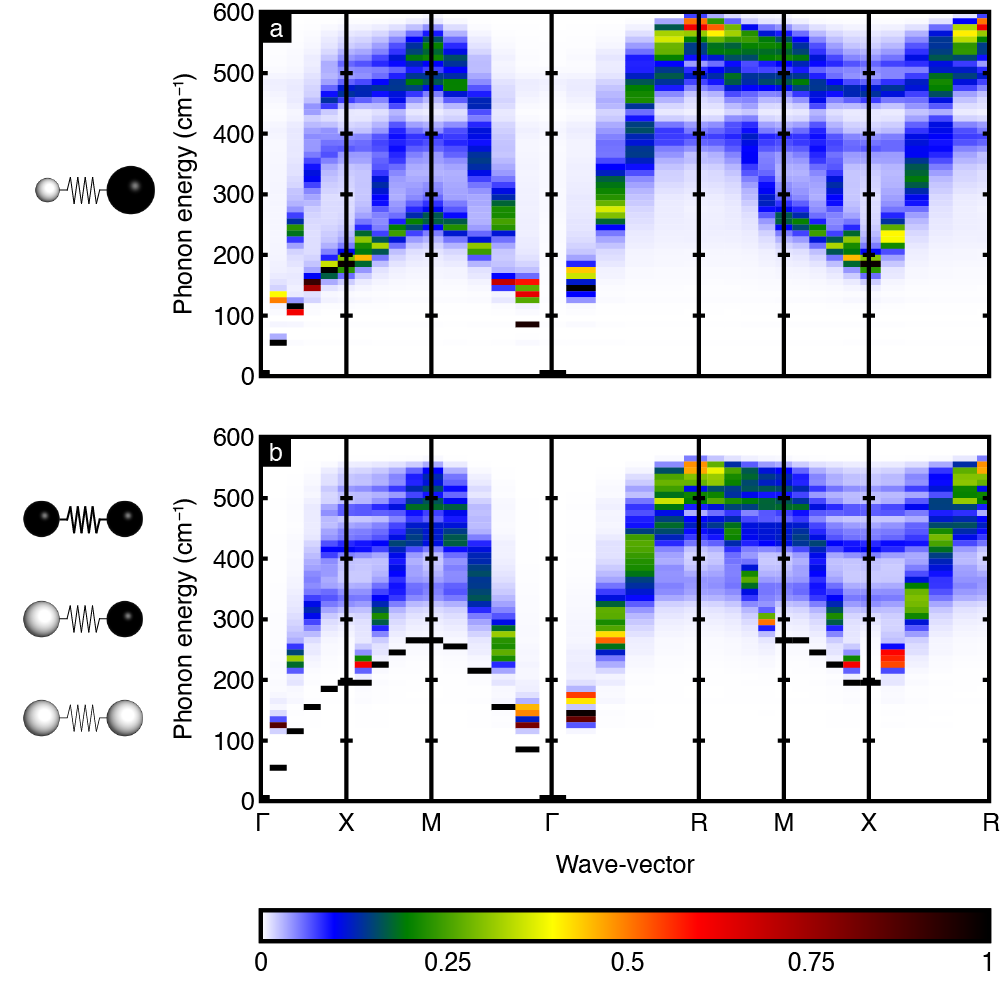}
\end{center}
\caption{A study into the effect of mass and $k_{\rm{harm}}$ disorder in a random binary alloy. (a,b) Phonon dispersion curves determined through the SCLD approach using a $10 \times 10 \times 10$ supercell for (a) a mass disordered alloy and (b) an alloy with harmonic bond stretching force constant disorder. \label{fig2}}
\end{figure}

In our first case study, we consider the effect of random mass disorder. This is implemented by varying $m_{\rm{A}}$ and $m_{\rm{B}}$ parameters but retaining homogeneity in the $k_{\rm{harm}},k_{\rm{angle}}$ terms [Table~\ref{table2}]. For a set of five independent configurations, each corresponding to a $10\times10\times10$ supercell of the simple cubic cell, we assigned atom sites to A and B components such that each configuration contained equal numbers of A and B atoms. The corresponding phonon spectrum, determined using the SCLD approach and averaged across all five configurations, is shown in Fig.~\ref{fig2}(a). An immediate consequence of the presence of nontrivial disorder is the continuous behaviour of the projection $\rho(\mathbf k,\omega)$, which now assumes non-integral values between 0 and 1. This spread in values of $\rho$ reflects the intrinsic contribution to phonon broadening arising purely due to static disorder within the configuration. Moreover additional branches appear to emerge (\textit{e.g.} at the R point around 400 cm$^{-1}$), showing that the SCLD approach, unlike perturbation theory, is able to capture multiple effects simultaneously. Qualitatively similar behaviour is observed for all ratios $m_{\rm{A}}/m_{\rm{B}}$, with the magnitude of phonon broadening a function of the deviation of this ratio away from unity. In comparing to experimental data one would need to take into account also the instrumental resolution function. 

\begin{table}[b]
\caption{The perturbed parameters in the mass and harmonic bond stiffness disordered alloys \label{table2}}
\begin{center}
\begin{tabular}{l|c|c}
&Mass&Force-constant\\
Parameter&disorder&disorder\\
\hline\hline
 $m_{\rm A}$ / amu & 40 & 60 \\
  $m_{\rm B}$ / amu & 80 & 60\\
   $k_{\rm{harm}}$  (A--A) / eV\AA $^{-2}$ & 10 & 5 \\
   $k_{\rm{harm}}$  (B--B) / eV\AA $^{-2}$ & 10 & 15\\
   $k_{\rm{harm}}$  (A--B) / eV\AA $^{-2}$ & 10 & 10\\
    \hline\hline
\end{tabular}
\end{center}
\end{table}

We performed similar calculations for an equivalent set of configurations but where we varied the strength of harmonic bond stretching force constants rather than atomic masses. The relevant parameters are again listed in Table~\ref{table2} and the corresponding phonon dispersion relations are now shown in Fig.~\ref{fig2}(b). The choice in parameters reflects a stronger effective variation than in the mass disorder and while the overall extent of phonon broadening is roughly similar to that observed for the case of mass disorder, we find a number of qualitative differences that demonstrate the sensitivity of the dynamical matrix to different types of disorder. For example, whereas the phonon density of states vanishes near 450\,cm$^{-1}$ in the case of mass disorder, the excitation spectrum of the system with force-constant disorder is entirely gapless. Perhaps even more striking is the absence of any broadening for specific branches of the phonon dispersion in Fig.~\ref{fig2}(b)---namely, the acoustic branches for $\Gamma$--X, X--M, and $\Gamma$--M directions in reciprocal space. That these two types of disorder give rise to different broadening effects is unsurprising since they effect diagonal and off-diagonal elements of the dynamical matrix in different ways \cite{Alam_2011}. Our results for these canonical systems agree well with the conclusions of earlier studies \cite{Larkin_2013, Beltukov_2013}.

\subsection*{Correlated disorder}

Our next step---and the ultimate goal of our study---was to explore the effect of including correlations in the distribution of A and B sites (while maintaining stoichiometry). We consider here the very simplest type of occupational correlation, corresponding to avoidance of like constituents on neighbouring sites; in zeolite chemistry this is the Lowenstein rule \cite{Lowenstein_1954}. The extent of correlation is quantified by the order parameter
\begin{equation}
\phi=\frac{N_{\rm{AB}}-(N_{\rm{AA}}+N_{\rm{BB}})}{3N},
\end{equation}
where $N_{\rm{AB}}$ is the number of nearest neighbour contacts between atoms of type A and B, and $N$ is the total number of atoms (= supercell size, in this case). A statistical distribution of A and B sites gives $N_{\rm{AA}}=N_{\rm{BB}}=3N/4$, $N_{\rm{AB}}=3N/2$ and hence $\phi=0$. If A--A/B--B avoidance is complete then $N_{\rm{AA}}=N_{\rm{BB}}=0$, $N_{\rm{AB}}=3N$ and $\phi=1$; this state corresponds to rocksalt-type ordering of A and B sites.

Using a custom Monte Carlo algorithm, we generated an ensemble of configurations corresponding to the order parameter values $\phi=0,\frac{1}{4},\frac{1}{2},\frac{3}{4},1$. Each configuration corresponded to a $12 \times 12 \times 12$ supercell of the primitive cubic cell. In almost every case we prepared five independent configurations to improve counting statistics; the exception was the case $\phi=1$, which is ordered. For each configuration, we calculated the corresponding phonon dispersion relation using our SCLD approach, averaging over configurations corresponding to equivalent values of the order parameter. For these various calculations, we used the `mass disorder' parameters given in Table~\ref{table2}. Consequently we expect the form of the phonon dispersion for the $\phi=0$ configurations to resemble closely that determined in the `random disorder' section above [Fig.~\ref{fig2}(a)]. 

\begin{figure}
\begin{center}
\includegraphics[width=\columnwidth]{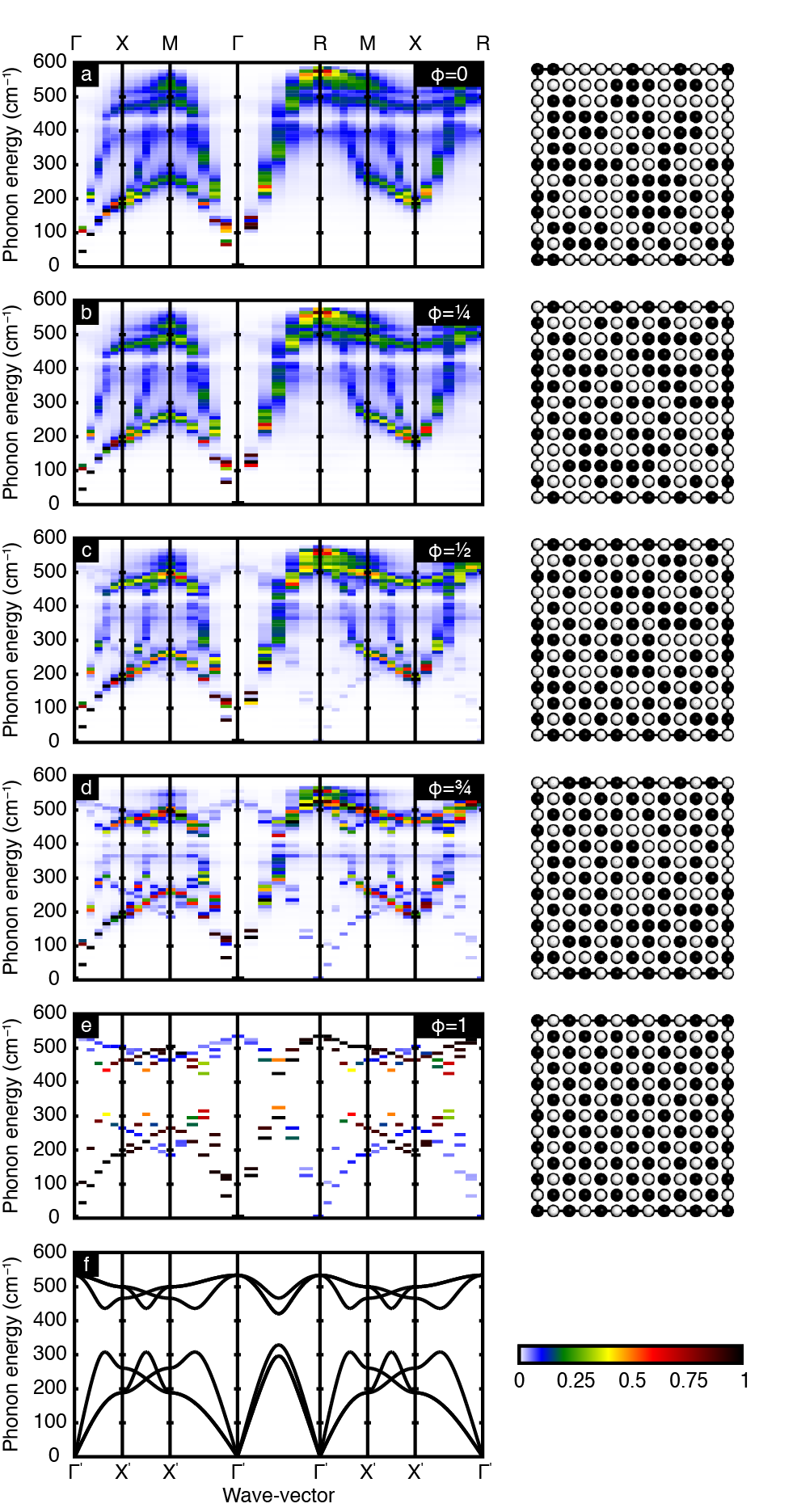}
\end{center}
\caption{Investigation into the effect of correlated disorder on phonon band broadening. (a-e) Phonon dispersion curves determined through the SCLD method for $12\times12\times12$ supercells of a mass disordered alloy with $\phi$ values of (a) 0.00, (b) 0.25, (c) 0.5, (d) 0.75 and (e) 1.00 -- the ordered NaCl structure. (f) Phonon dispersion curves determined using the conventional dynamical matrix calculations in GULP for the $\phi =1.00$ structure. The differing wave-vector axis labels at the top and bottom of the figure refer to the high symmetry points in the primitive cubic and face centred cubic Brillouin zones respectively. The colour scale is a measure of the projection $\rho$ at each $\mathbf{k}$-point. \label{fig3}}
\end{figure}

The resulting dispersion curves are shown in Fig.~\ref{fig3}. We do indeed observe good consistency between our results for $\phi=0$ [Fig.~\ref{fig3}(a)] and those given in Fig.~\ref{fig2}, noting for completeness that the $\mathbf k$-grid is slightly denser for the former than for the latter. Our key result, however, is the consistent and physically sensible evolution of the phonon spectrum as the strength of correlations increases. We find a steady decrease in phonon broadening with increasing $\phi$ that accompanies the opening of an excitation gap near 400\,cm$^{-1}$. Hence, in a real system, the nature of the density of states in this region of the phonon spectrum is likely to be highly sensitive to the degree of site ordering. Also clear is the emergence of a new set of phonon branches that in the case of $\phi=1$ establishes the original zone corner $\mathbf k=(\frac{1}{2},\frac{1}{2},\frac{1}{2})$ (R) as a new zone centre (denoted in Fig.~\ref{fig3} as $\Gamma^\prime$). This behaviour is entirely physical: the ordered $\phi=1$ configuration has a face-centred cubic unit cell with lattice parameter double that of the original primitive cubic cell. Hence the point $\mathbf k=(\frac{1}{2},\frac{1}{2},\frac{1}{2})$ is now a reciprocal lattice vector. The harmonic phonon relation can be calculated exactly for this state and is shown in Fig.~\ref{fig3}(f); the match to our SCLD result in Fig.~\ref{fig3}(e) is excellent.

We note that the projection magnitude $\rho(\mathbf k,\omega)$ for the secondary (face-centred) phonon branches is much smaller than that for the primary branches. This difference reflects the mass difference between A and B atoms. It is straightforward to show, using standard lattice dynamical theory \cite{Dove_1993}, that the projections of the acoustic and optic branches at the zone centre are given by
\begin{eqnarray}
\rho(\mathbf k=\mathbf 0,{\rm{acoustic}})&=&\frac{1}{2}+\frac{\sqrt{m_{\rm{A}}m_{\rm{B}}}}{m_{\rm{A}}+m_{\rm{B}}},\\
\rho(\mathbf k=\mathbf 0,{\rm{optic}})&=&\frac{1}{2}-\frac{\sqrt{m_{\rm{A}}m_{\rm{B}}}}{m_{\rm{A}}+m_{\rm{B}}}.
\end{eqnarray}
Hence, in the limit of large mass difference between A and B atoms, the projections approach $\rho\rightarrow\frac{1}{2}$; conversely, for vanishingly small mass differences, the `optic' component also vanishes and the phonon dispersion reverts to that observed for a single component system. In our particular case, the values of $m_{\rm{A}}$ and $m_{\rm{B}}$ given in Table~\ref{table2} correspond to projection values of 97\% (acoustic) and 3\% (optic), as shown in Fig.~\ref{fig3}(e).

While we have focussed here on an extremely simple case of correlated disorder, the approach we have developed is straightforwardly applied to substantially more complicated cases. Our particular eventual interest is in strongly-correlated systems such as the various `procrystalline' families described in Ref.~\citenum{Overy_2016}. We note that these systems include examples where the source of disorder is electronic in nature, such as the correlated second-order Jahn Teller displacements observed in cubic BaTiO$_3$\cite{Senn_2016,Comes_1970}. We anticipate that application of SCLD methods to such systems may help bridge the ostensibly unrelated concepts of disorder- and anharmonicity-dependent phonon broadening.

\section{Concluding Remarks}

Access to computationally tractable calculation approaches for determining phonon relations in disordered materials opens up the attractive possibility of iterative comparison with experimental inelastic neutron scattering and/or inelastic X-ray scattering measurements. This offers a means of refining both the form of structural disorder and the perturbations of effective interaction parameters from the experimental phonon spectrum. Such an approach would complement conventional elastic scattering measurements as a method of characterising disorder \cite{Welberry_2016}, with the added advantage of providing insight into the effect of local symmetry breaking on local interaction strengths. Quantitative comparison with experimental data would require accurate characterisation of the experimental resolution function, with care taken also to ensure appropriate treatment of polarisation effects (\emph{e.g.}\ LO/TO splitting).

We conclude with two final points of potential relevance to subsequent applications of SCLD methods. First, the absence of explicit symmetry constraints in the SCLD approach allows in principle for the investigation of infrared/Raman selection rule violations in disordered materials, including canonical systems such as water ice \cite{Uwe_1986, Inoue_1988, Fukazawa_2000, Wong_1976}. And, second, we note that the phonon broadening we observe is of course a signature of the breakdown of the plane-wave description of vibrational states in these disordered solids. Exploration of the localisation (or otherwise) of the corresponding eigenvectors may provide valuable insight into the relationship of these vibrational states and the diffusons implicated in thermal transport in glasses \cite{Beltukov_2013, Prasai_2016}. 


\section*{Acknowledgements}

A.R.O., A.S., and A.L.G. gratefully acknowledge financial support from the European Research Council (Grant 279705), from the Diamond Light Source, UK to A.R.O., and to A.S. from the Leverhulme Trust (Grant RPG-2015-292) and the Swiss National Science Foundation (Grant P2EZP2\_155608).

\section*{References}






\end{document}